\documentclass{article}
\usepackage[T1]{fontenc}
\usepackage[latin1]{inputenc}
\usepackage{graphics}
\usepackage{epsfig}
\IfFileExists{url.sty}{\usepackage{url}}
                      {\newcommand{\url}{\texttt}}

\makeatletter

\providecommand{\LyX}{L\kern-.1667em\lower.25em\hbox{Y}\kern-.125emX\@}
\newcommand{\noun}[1]{\textsc{#1}}

 \newenvironment{lyxcode}
   {\begin{list}{}{
     \setlength{\rightmargin}{\leftmargin}
     \raggedright
     \setlength{\itemsep}{0pt}
     \setlength{\parsep}{0pt}
     \verbatim@font}%
    \item[]}
   {\end{list}}
 \newcommand{\lyxaddress}[1]{
   \par {\raggedright #1 
   \vspace{1.4em}
   \noindent\par}
 }

\def\lapproxeq{\lower .7ex\hbox{$\;\stackrel{\textstyle<}{\sim}\;$}}
\def\gapproxeq{\lower .7ex\hbox{$\;\stackrel{\textstyle>}{\sim}\;$}}
\newcommand{\xpom}{x_{_{\rm I\!P}}}
\newcommand{\pom}{{\rm I\!P}}
\newcommand{\regge}{{\rm I\!R}}
\makeatother

\begin{document}

\title{Diffractive vector boson production at the Tevatron}

\author{B.E. Cox$^a$, J.R. Forshaw$^a$ and L. L\"onnblad$^b$}

\maketitle

\lyxaddress{$^a$Department of Physics \& Astronomy, University of
  Manchester. Manchester.  M13 9PL. UK. }

\lyxaddress{$^b$Department of Theoretical Physics 2, S\"olvegatan 14A,
  S-223 62 Lund. Sweden.}

\begin{abstract}
  The diffractive production of vector bosons at the Tevatron is
  studied. We take a hard pomeron flux and use the H1 parton density
  functions extracted from their measurement of $F_2^{D(3)}$. To this
  we add a reggeon exchange contribution.  We find that the ratio of
  diffractive to non-diffractive $W$ boson production is in good
  agreement with the \noun{CDF} data. We note that the poorly
  understood reggeon contribution might be as large as the pomeron
  contribution in the kinematic range of the data. We also note that
  our pomeron exchange contribution is much smaller than earlier
  predictions due to our use of a hard pomeron flux. All our results
  have been obtained using the lowest order matrix elements and we
  have compared to the results obtained using a modified version of
  \noun{Herwig} which is capable of generating a wide variety of
  diffractive scattering processes.  Gap survival is estimated using
  \noun{Pythia} and is found to be around 60\%.  We make predictions
  which could be compared to future measurements of diffractive
  Drell-Yan production.
\end{abstract}
\begin{lyxcode}
\vspace*{-17cm}

\begin{flushright}

LU~TP~00-57~\\
MAN/HEP/2000/4~\\
MC/TH/00/12\\
hep-ph/0012310

\end{flushright}

\newpage
\end{lyxcode}

\section{Introduction}

So far there is no consensus on the interpretation of the hard
diffractive scattering processes measured at the \noun{Tevatron
  \cite{SSjets,diffw,DPE,diffb}.}  Invariably the rates measured at
the \noun{Tevatron} are much smaller than would be expected using the
diffractive parton density functions measured in diffractive deep
inelastic scattering at \noun{Hera} \cite{H1,ZEUS}. On the one hand
this is not in contradiction with theory since the diffractive parton
density functions extracted in deep inelastic scattering are not
expected to be applicable in diffractive scattering involving incoming
hadrons \cite{Collins}.  However, there is the hope that this breaking
of universality might be described in terms of an overall
multiplicative ``gap survival'' factor \cite{gap-survival}.

In this paper we turn our attention to diffractive vector boson
production at the \noun{Tevatron}. We perform an analysis using the
diffractive parton density functions extracted by the H1 collaboration
\cite{H1} with one important modification.  Namely, we use a larger
effective pomeron intercept, i.e. $\alpha_{\pom}(0) = 1.45$
rather than the value 1.203 extracted by H1\footnote{%
  We speak of the ``effective pomeron intercept'' since it is
  commonplace. However, what we refer to is the parameter which
  controls the $\xpom$ distribution, assuming a factorisation of the
  diffractive parton density into a product of a function of $\xpom$
  and a function of $\beta$.  }. We believe this to be a natural
choice given that the scale associated with diffractive $W$ production
is some two orders of magnitude larger than that in diffractive
\noun{dis} at \noun{Hera,} and given that there is evidence for an
effective pomeron intercept which increases slowly as the hard scale
increases \cite{rise,DL-2}. We include the rather poorly constrained
reggeon exchange contribution, also as determined by the H1
\noun{}collaboration \cite{H1-98}.

In Section \ref{sec:W} we focus on $W$ boson production as measured by
\noun{cdf \cite{diffw}.} This process has been studied within a
similar framework previously \cite{Alvero,Covolan} where it was found
that the rates based upon direct extrapolation of the deep inelastic
parton density functions typically overestimate the \noun{cdf} data by
an order of magnitude. However, neither of these analyses used a hard
pomeron flux, nor did they consider the reggeon contribution. Possible
explanations for the largeness of the theoretical cross-section have
been presented \cite{Covolan,Dino-Tan}, some based upon the concept of
gap survival \cite{Levin,Martin}. We shall discuss the issue of gap
survival in Section \ref{sec:GS}.

Throughout our study, we shall perform our calculations using the
relevant leading order electroweak matrix elements. In addition, we
compare to results obtained using a modified version of \noun{Herwig,}
which we call \noun{Pomwig,} that is able to generate diffractive
scattering in a wide range of processes\cite{Herwig,Pomwig}.\footnote{%
  The code can be obtained from the authors upon request or from\\
  \url{www.hep.man.ac.uk/pomwig/} .  }

Our studies of diffractive $W$ production conclude that the reggeon
contribution is probably not negligible. It is poorly understood and
more studies are needed in order to reduce the uncertainty from this
source. Furthermore, we propose that the rate for pomeron initiated
processes is much smaller than has been found hitherto, due mainly to
our use of diffractive parton densities with a steeper $\xpom$
dependence than occurs in softer pomeron exchange models. In Section
\ref{sec:DY} we examine diffractive Drell-Yan production in order to
investigate the additional information that can be obtained at the
\noun{Tevatron}.  In Section \ref{sec:conclusions} we present our
conclusions.

\section{Diffractive $W$ production\label{sec:W}}

For diffractive vector boson production in proton-antiproton
collisions with a fast forward proton we use the general formula\\
\begin{equation}
  \frac{d\sigma}{d\eta}= \frac{1}{2 s} \sum_{ij} \int
  d(\cos \theta) \int \frac{d\xpom}{\xpom}\int d\hat{s} \,
  F_{\pom/p}(\xpom) f_{i,\pom}(\beta,\mu^2) f_{j,\bar{p}}(x,\mu^2)
  \hat{s} \frac{d \hat{\sigma}_{ij}}{d \hat{t}}
\end{equation}
for the rapidity ($\eta$) distribution of the lepton which is produced
in the decay of the boson. Using the language of Ingelman \& Schlein
\cite{IS}, the pomeron flux factor is $F_{\pom/p}(\xpom)$ and
$f_{i,\pom}(\beta,\mu^2)$ is the parton distribution function for
partons of flavour $i$ in the pomeron.  As is conventional, $\xpom$ is
the fraction of the proton energy carried by the pomeron and $\beta$
is the fraction of the pomeron energy carried by the parton involved
in the hard scatter. The $p\bar{p}$ centre-of-mass energy is denoted
$s = (1800$ GeV$)^2$. $ f_{j,\bar{p}}(x,\mu^2)$ is the anti-proton
distribution for partons of flavour $j$. Throughout we take the
factorisation scale $\mu$ to be the subprocess centre-of-mass energy
$\hat{s}$.

In the particular case of $W^-$ production we use the narrow width
approximation to write the rapidity distribution of the scattered
electron\footnote{%
  For definiteness we speak of the 1st generation of leptons only.
  Including the 2nd generation will of course increase all our
  cross-sections by a factor of two. }
(throughout this paper positive rapidity is defined in the proton direction):\\
\begin{equation}
  \frac{d\sigma}{d\eta} = \frac{G_F^2 M_W}{12 s \Gamma_W} \sum_{ij} \int
  d(\cos \theta) \int \frac{d\xpom}{\xpom} \,
  F_{\pom/p}(\xpom) f_{i,\pom}(\beta,\mu^2) 
  f_{j,\bar{p}}(x,\mu^2) \ V_{ij} \ \hat{Q}_{ij}^4 
\end{equation}
$V_{ij} = \cos^2 \theta_c$ for $\bar{u}d$ and $\bar{c}s$ incoming
partons or $V_{ij} = \sin^2 \theta_c$ for $\bar{u}s$ and $\bar{c}d$
incoming partons. $\hat{Q}_{ij}^2 = \hat{u}$ in the case that a quark
from the pomeron participates in the hard scatter and
$\hat{Q}_{ij}^2 = \hat{t}$ when it is an anti-quark, where\\
$$ \hat{u} = -\frac{M_W^2}{2}(1+\cos \theta), $$
$$ \hat{t} = -\frac{M_W^2}{2}(1-\cos \theta), $$
and $\theta$ is the scattering angle of the scattered electron in the
parton-parton centre-of-mass frame measured relative to the proton
direction. In the case of $W^+$ production it is necessary to make the
interchange $\hat{t} \leftrightarrow \hat{u}$ if we are to define
$\eta$ as the rapidity of the scattered positron. We integrate over
$|\cos \theta| < ( 1-4p_{t{\rm min}}^2/M_W^2)^{1/2}$ with $p_{t{\rm
    min}} = 20$ GeV and all $\xpom < 0.1$. These are the cuts used by
the \noun{cdf} collaboration.

In order to compute the ratio of diffractive to non-diffractive events
we use a similar expression for the non-diffractive cross-section,
i.e. by replacing the pomeron flux factor by $\delta(1-\xpom)$ and the
pomeron parton densities by the proton ones. In computing this ratio
we integrate over the \noun{cdf} range, $|\eta| < 1.1$.

Our default is to use the H1 pomeron flux, with an increased value for
the pomeron intercept, i.e.\\
\begin{equation}\\
  F_{\pom/p}(\xpom) =
  N \int dt \ \exp(4.6t) \left( \frac{1}{\xpom}\right)^{1.90+0.52t}
\end{equation}
and integrate over all $t$ (in practise it is sufficient to integrate
over $0 < -t < 1$ GeV$^2$). The normalisation $N$ is defined so that
the flux is equal to 1 at $\xpom = 0.003$.\footnote{%
  Using the routine \noun{qcd\_1994}\cite{h1pdf}.  } Note that this is
a significantly steeper flux than that proposed by Donnachie \&
Landshoff (\noun{dl})\cite{DL}, which is roughly equivalent to the
above flux except for the replacement of 1.90 by 1.17. For the pomeron
parton densities we use the favoured leading order H1 \noun{}fit 2. We
always use the \noun{GRV94 lo} parton densities for incoming hadrons
\cite{GRV}.

For the reggeon we simply replace the pomeron flux with \cite{pion-flux}\\
\begin{equation}\\
  F_{\regge/p}(\xpom) = \frac{390}{8 \pi^2}
  \int dt \ \exp(4.0t) \left( \frac{1}{\xpom}\right)^{0.50+0.90t} 
  \label{reggeon}
\end{equation}
and the pomeron parton densities are replaced by the \noun{grv}
leading order pion densities of \cite{GRV-pion}. This is the reggeon
contribution that was shown to agree with the H1 data collected in the
range $0.1 < \xpom < 0.3$ \cite{H1-98}. We ignore any possible
interference between the pomeron and the reggeon although this may be
important.

\begin{figure}[t]
  \begin{center}
    \parbox{16cm}{\hspace*{-0.5cm}
      \epsfig{figure=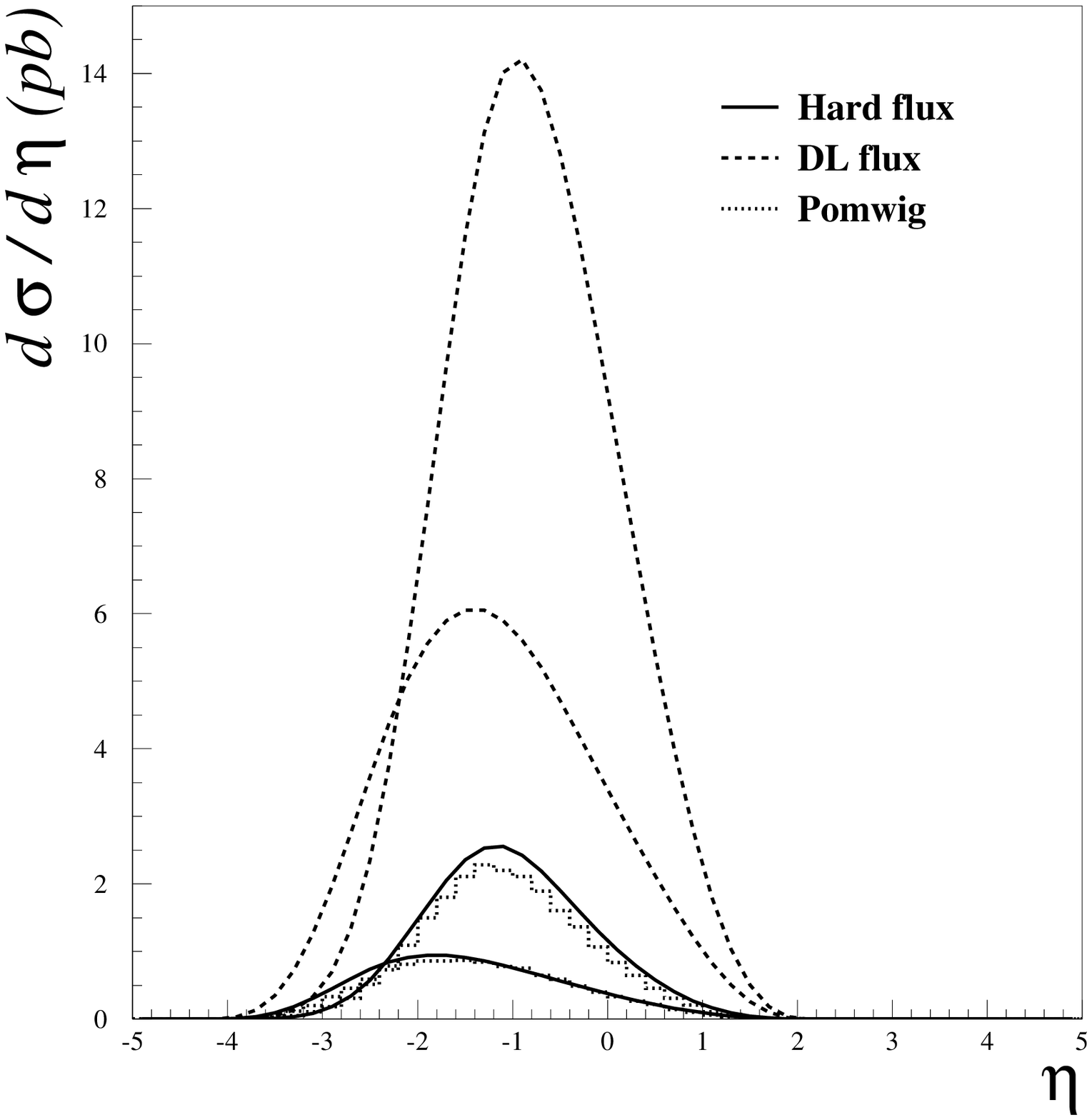,height=7cm,width=7cm}\hspace*{-1cm}
      \epsfig{figure=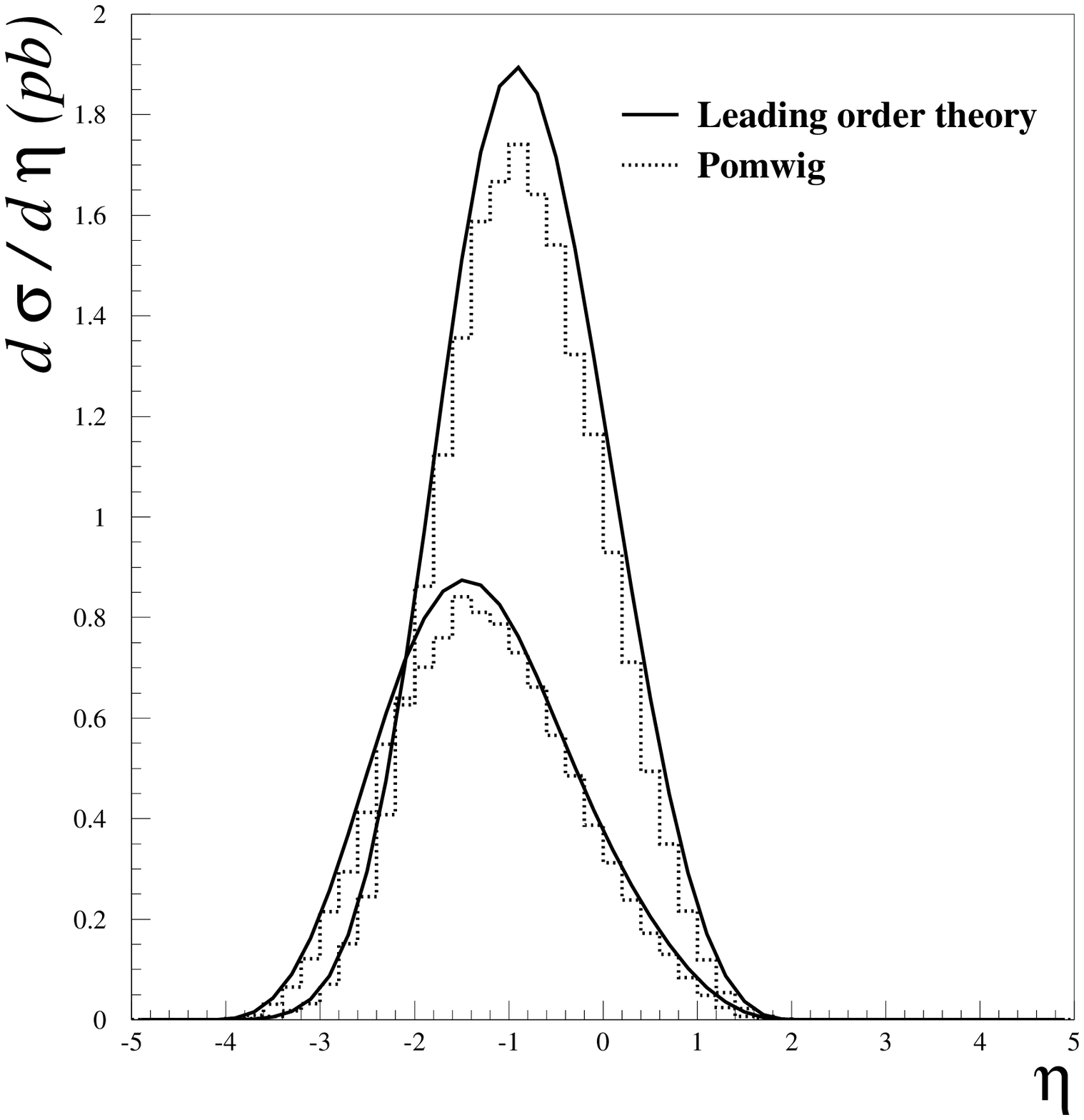,height=7cm,width=7cm}}
  \end{center}
  \caption{\label{Fig:W-theory}
    The rapidity distribution of the electron and positron for the
    pomeron (upper plot) and reggeon (lower plot) contributions in
    diffractive $W$ production. In all cases the electron distribution
    is peaked at more positive rapidity than the positron. In the
    pomeron case, the broken lines correspond to using the flux of
    Donnachie \& Landshoff rather than our default flux. The histogram
    is the result from \noun{Pomwig.}}
\end{figure}

In Figure \ref{Fig:W-theory} we show the rapidity distribution of the
scattered lepton for both the pomeron and reggeon contributions. In
all cases the electron distribution is more forward peaked than the
positron due to the chiral nature of the $W$ interactions. Note the
significant suppression obtained when using the hard flux as compared
to the \noun{dl} flux. This is a consequence of the flatter $\xpom$
distribution in the latter case -- a small effect at $\xpom$ values
typical for \noun{Hera} but not so for the much larger $\xpom$ values
probed at the \noun{Tevatron}. The fact that the \noun{Tevatron} is at
significantly larger $\xpom$ is also responsible for the large
contribution from the reggeon.  Also significant is the fact that the
peaks of the lepton distributions are more backward when using the
hard flux compared to the \noun{dl} flux. This arises because the
larger $\xpom$ values in the \noun{dl} case lead to a smaller boost to
negative rapidity than occurs using the H1 flux. The implication is
that when we integrate over the \noun{cdf} acceptance region ($|\eta|
< 1.1$) a larger fraction of the leptons produced using the hard flux
fall outside the region of acceptance thereby still further reducing
the predicted cross-section.

The bottom line is that the ratio of diffractive to non-diffractive
events, summed over the electron and positron, and multiplied by 2 to
account for proton and antiproton dissociation, is much smaller in the
case of the \noun{}hard flux compared to the \noun{dl} flux. In
particular we find $R_{{\rm DL}}^{\pom} = 6.38\%$ and $R_{{\rm
    hard}}^{\pom} = 0.86\%$ where $R^{\pom}$ denotes the ratio of
pomeron exchange events to non-diffractive events integrated over the
\noun{cdf} acceptance region. This is to be compared to the \noun{cdf}
value for the ratio of diffractive to non-diffractive events, $R_{{\rm
    CDF}} = 1.15 \pm 0.55 \% $. However, we cannot avoid the reggeon
contribution. Including it leads to a ratio $$R^{\pom+\regge}_{{\rm
    hard}} = 1.67\%.$$
Using \noun{Pomwig,} this ratio is only very
slightly different at 1.72\%.

Before leaving this section it is necessary to make a few remarks
regarding the uncertainty in this theoretical result. There is a large
uncertainty associated with what we have called the effective pomeron
intercept, since there is no existing measurement of this parameter in
the kinematic range relevant to the T\noun{evatron}. We have used the
\noun{H1} fits for the diffractive parton density functions. These
density functions are only evolved up to $Q^2 = 75$ GeV$^2$ and remain
constant thereafter. We have also obtained results using the
\noun{actw} parton density functions \cite{Alvero}, which are evolved
to high $Q^2$ and from which we are able to estimate that normal
\noun{qcd} evolution will change our prediction for the gap fraction
by no more than 20\%. Next-to-leading order \noun{qcd} corrections are
expected to play a role at the 10-20\% level. There is also an
uncertainty in the size of the reggeon contribution that is hard to
quantify.

\section{Gap survival} \label{sec:GS}

So far we have ignored the fact that factorisation is not expected to
hold in hard diffraction at the \noun{Tevatron}. One might attempt to
account for this physics by an overall multiplicative factor which
specifies the probability that the rapidity gap is destroyed by
additional interactions which are not strongly correlated with the
primary hard scatter. We aim in this section to add plausibility to
this assumption and to make an estimate for the gap survival factor.

In a previous paper \cite{CFL} we used the multiple interactions model
implemented in the \noun{Pythia} event generator to estimate the gap
survival probability in gap-between jets events
\cite{Pythia,Pythia-MI}. In this model the number of additional
scatters, given a particular hard trigger process, is dependent only
on the total energy of the collisions and of the scale of the hard
trigger process. The estimate of the survival probability obtained was
22\% for 1800 GeV and 35\% for 630 GeV and was largely independent of
the transverse momentum and rapidity of the jets.

The increase of multiple interactions with the total energy is simply
due to the small-$x$ proliferation of gluons which are available for
additional scatterings.  The dependence on the hard scale can be
described in a simplified manner as follows.

The model assumes that the partons in the proton are distributed
according to a double Gaussian in impact-parameter space. This means
that central collisions with a small impact parameter have a higher
probability for multiple scatterers than do peripheral ones with a
large impact parameter. For the case of $W$-production, which we are
interested in here, the procedure is to generate the hard electroweak
scattering and then to choose an impact parameter depending on the
scale of the process -- the higher the scale the larger the
probability to pick a small impact parameter. The overlap is then
calculated from this and the number of additional scatterers is
obtained. (Note that the procedure becomes more complicated if a
trigger process with large cross section is used, since the
eikonalisation of the cross section becomes important in that case.)

In principle it is possible to run \noun{Pompyt} \cite{pompyt} with
the multiple interactions of \noun{Pythia}. This will, however, only
give multiple interactions between the colliding anti-proton and
pomeron, and such scatterings will not destroy the gap. What we need
is to estimate the multiple interactions between the proton and
anti-proton. Here the scale which determines the overlap is not the
hard scale of the electroweak $W$-production process, but rather the
momentum transfer of the pomeron $-t$, which is small. So, although
the overlap between the pomeron and anti-proton is large, the overlap
between the proton and anti-proton may be small, resulting in a
gap-survival probability much larger than the ones quoted above.
\begin{figure}
  {\par\centering \resizebox*{0.7\textwidth}{!}{\rotatebox{-90}
      {\includegraphics{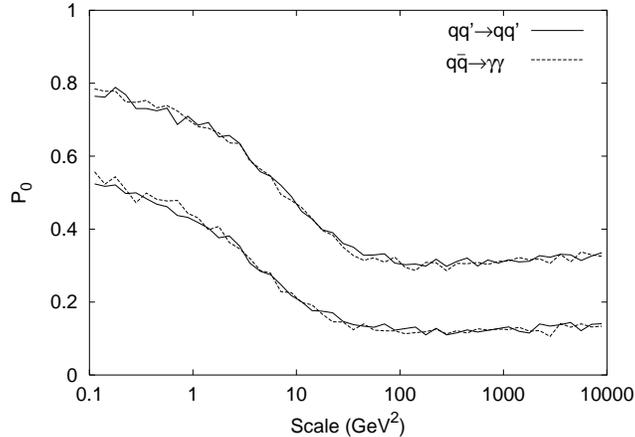}}} \par}

  \caption{\label{Fig:GS}
    The fraction of events without additional scatterings ($P_0$) as a
    function of the hard scale in photon exchange (full lines) and
    di-photon production (dashed lines) events. The upper curves are
    the fraction of events with no additional colourful scatterings
    whilst the lower ones are the fractions with no additional
    scatterings at all.}
\end{figure}

Since it is not possible simply to add multiple interactions between
the proton and anti-proton in \noun{Pompyt}, we use the fact that the
amount of multiple scattering only depends upon the scale of the
process and the total collision energy. Using different triggering
processes which are able to go to small scales without large
eikonalisation effects in \noun{Pythia} (using default parameters) we
find that the probability not to have any additional scatters
saturates for small momentum transfers at around 50\%. In Figure
\ref{Fig:GS} we show this probability for proton-antiproton scattering
via the electromagnetic $qq' \to qq'$ and $q \bar{q} \to \gamma
\gamma$ sub-processes.

For large scales the probability is well below the 22\% we quoted in
\cite{CFL}.  This is because of the way partons from multiple
scatterings are colour-connected to the rest of the event in
\noun{Pythia.} With the default parameters only 1/3 of the multiple
scatterings will be colour-connected to the rest of the event, while
the remaining scatterings will produce colour singlet $q \bar{q}$ or
$gg$-pairs which will not necessarily destroy the gap. In Figure
\ref{Fig:GS} we therefore also show the probability of having at least
one colourful secondary scatter. The true gap fraction ought to lie
somewhere in between these two extremes.

There are large uncertainties here. Multiple interactions are poorly
understood and the multiple interaction model in \noun{Pythia} has not
been tuned to describe these processes. It is, however, natural to
assume that any rapidity gap produced due to the exchange of a pomeron
or a reggeon may be destroyed by multiple interactions. It is also
very intuitive that these processes are more peripheral than other
hard interactions and that the survival probability therefore should
be larger than in the case of, e.g. high-$t$ diffraction. The model in
\noun{Pythia} does attempt to take such physics into account.
Nevertheless, it is clear that more effort must be devoted to
understanding these processes before we can start to compare the
diffractive results from \noun{Hera} with the ones from the
\noun{Tevatron} with confidence.

The results of this section suggest that we should take a gap survival
factor of around 60\% for low $t$ diffraction at the \noun{Tevatron,}
running at 1.8 TeV. This should be compared to other models which
typically predict gap survival factors of 10\% or less
\cite{Levin,Martin}. Accounting for this gap survival factor we find a
gap fraction of

$$ R^{\pom+\regge}_{\rm hard} = 1.00\% $$
which is in good agreement with the \noun{Tevatron} measurement.

\section{Diffractive Drell-Yan production\label{sec:DY} }

In order to test our understanding of diffractive vector particle
production, it would be useful not only to measure the rapidity
distribution of the charged lepton in $W^{\pm}$ production but also to
do the same for the diffractive Drell-Yan production process and in
this section we present our predictions.
\begin{figure}
  {\par\centering \resizebox*{0.7\textwidth}{!}{\includegraphics{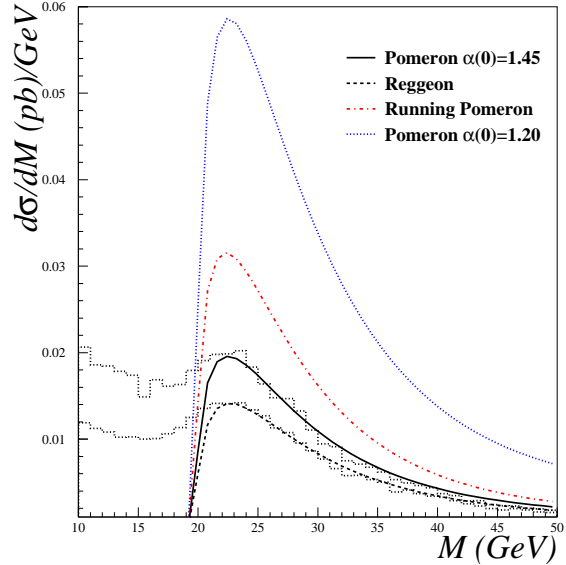}} \par}

  \caption{\label{Fig:DY-theory}
    The cross-section as a function of invariant mass for the
    Drell-Yan process. The solid line is the pomeron contribution and
    the dashed line is the reggeon contribution. The curves are
    obtained for $p_T > 10$ GeV.  The dotted line is the pomeron
    contribution for a smaller effective intercept and the dot-dash
    line shows the same but with a variable effective intercept, as
    explained in the text. The histrograms are the \noun{Pomwig}
    result.}
\end{figure}

\newpage
The rapidity distribution of the scattered lepton can be computed:\\
\begin{eqnarray}
  \frac{d\sigma}{d\eta}&=& \frac{1}{32 \pi s} \sum_{ij} \int
  d(\cos \theta) \int \frac{d\xpom}{\xpom}\int \frac{d\hat{s}}{\hat{s}} \,
  F_{\pom/p}(\xpom) f_{i,\pom}(\beta,\mu^2) f_{j,\bar{p}}(x,\mu^2) \nonumber \\
  &\times & \frac{2 e^4 e_q^2}{3}
  \left(\frac{\hat{u}^2 + \hat{t}^2}{\hat{s}^2}\right).
\end{eqnarray}
The advantage of the Drell-Yan process is the ability to tune the
invariant mass of the dilepton pair and in Figure \ref{Fig:DY-theory}
we show the cross-section as a function of invariant mass, $M \equiv
\sqrt{\hat{s}}$, integrated over $-1.1 < \eta < 1.1$ for both the
pomeron and reggeon contributions. The solid black line represents a
hard pomeron with effective intercept $\alpha_{\pom}(0) = 1.45$, the
dotted line represents a hard pomeron with an effective intercept
$\alpha_{\pom}(0) = 1.20$, and the broken line is the reggeon
contribution. The histrograms are the \noun{Pomwig} results, and the
non-vanishing at low $M$ is a result of the correct matching to the
\noun{nlo} matrix elements that is performed in \noun{Herwig} 6.1.

To match smoothly with the \noun{Hera} data it is perhaps more
reasonable to use an effective intercept that depends upon the hard
subprocess scale. This possibility leads to the dash-dot curve in
Figure \ref{Fig:DY-theory} where we have used
$$ \alpha_{\pom}(0) = 1.10 + 0.038 \ln M^2.$$
We wish to stress that this is an entirely phenomenological
parameterisation. From a theoretical perspective, such an $\xpom$
dependence cannot be generated by exchange of a single regge pole.
However this effective behaviour could describe the interplay of two
or more regge exchanges. For example, in a two-pomeron model one might
imagine that the harder pomeron, with intercept around 1.45, dominates
at the \noun{Tevatron} whilst at \noun{Hera} there is an important
admixture of the softer pomeron - this would lead to qualitatively the
behaviour we examine here \cite{DL-2}. Alternatively, within the
leading-twist \noun{pqcd} formalism, the $\xpom$-dependence should not
exhibit any $Q^2$ dependence since all of the evolution is contained
within what we call the pomeron parton distribution functions. In this
case, one should interpret our parameterization as an indication that
there is an important higher-twist component in the relatively low
$Q^2$ \noun{Hera} data that is absent in the \noun{Tevatron} data.

\section{Conclusions \& Outlook\label{sec:conclusions}}

We have shown that the current \noun{Tevatron} data on diffractive
$W^{\pm}$ production can be understood in terms of the diffractive
parton densities extracted from \noun{Hera} data provided we use a
hard pomeron intercept, $\alpha_\pom(0) = 1.45$.  Modelling the
diffractive parton densities in terms of pomeron and reggeon
contributions, we find that the reggeon contribution is approximately
equal to the pomeron contribution in the kinematic regime relevant for
the \noun{Tevatron.} Moreover, we propose that the pomeron
contribution is much smaller than previous estimates.  We used
\noun{Pythia} to model the gap survival probability and found it to be
quite large at 60\%. In conclusion we find a gap fraction in $W^{\pm}$
production of 1.00\% which is to be compared to the \noun{cdf} result
of $(1.15 \pm 0.55)\%$.

A precise comparison between theory and data is not yet possible due
primarily to lack of understanding of gap survival effects and of the
reggeon contribution.  Future data collected on diffractive Drell-Yan
production at the \noun{Tevatron} will certainly help clarify the
physics, not least by providing valuable data over a wide range of
scales, from those typical of \noun{dis} at H\noun{era} to those in
diffractive $Z^0$ production\noun{.} Measurements to lower $\xpom$ at
the \noun{Tevatron} will also be important in helping disentangle the
competing effects.

The natural next step is to compare the model presented here to the
data for diffractive dijet production \cite{SSjets} and diffractive
central jet production \cite{DPE}. Preliminary investigations suggest
that the existing contradictions will be resolved - we deal with these
processes in a future publication.

\section*{Acknowledgements}

We should like to thank Dino Goulianos, Hannes Jung, Andy Mehta, Paul
Newman, Julian Phillips and Mike Seymour for their valuable
assistance.


\begin{thebibliography}{10}
\bibitem{SSjets}
  F.~Abe et al., CDF Collaboration, Phys.\ Rev.\ Lett.\ 79, 2636 (1997); \\
  T.~Affolder et al., CDF Collaboration, Phys.\ Rev.\ Lett.\ 84, 5043 (2000)
\bibitem{diffw}
  F.~Abe et al., CDF Collaboration, Phys.\ Rev.\ Lett.\ 78 (1997) 2698.
\bibitem{DPE}
  T.~Affolder et al., CDF Collaboration, Phys.\ Rev.\ Lett.\ 85, 4215 (2000).
\bibitem{diffb}
  T.~Affolder et al., CDF Collaboration, Phys.\ Rev.\ Lett.\ 84, 232 (2000).
\bibitem{H1}
  T.~Ahmed et al., H1 Collaboration, Phys.\ Lett.\ B348 (1995) 681; \\
  C.~Adloff et al., H1 Collaboration, Z.~Phys.\ C76 (1997) 613.
\bibitem{ZEUS}
  J.~Breitweg et al., ZEUS Collaboration, Eur.\ Phys.~J.\ C1 (1998) 81; \\
  Eur.\ Phys.~J.\ C6 (1999) 43.
\bibitem{Collins}
  J.C.~Collins, Phys.\ Rev.\ D57 (1998) 3051.
\bibitem{gap-survival}
  J.D.~Bjorken, Phys.\ Rev.\ D47 (1993) 10.
\bibitem{rise}
  For example see B.E.~Cox, J.~Phys.\ G25 (1999) 1377.
\bibitem{DL-2}
  A.~Donnachie and P.V.~Landshoff, Phys.\ Lett.\ B437 (1998) 408.
\bibitem{H1-98}
  C.~Adloff et al., H1 Collaboration, Eur.\ Phys.~J.\ C6 (1999) 587.
\bibitem{Alvero}
  L.~Alvero, et al., Phys.\ Rev.\ D59 (1999) 074022 
\bibitem{Covolan}
  R.J.M.~Covolan and M.S.~Soares, Phys.\ Rev.\ D60 (1999) 054005;\\
  erratum-ibid. D61 (2000) 01990.
\bibitem{Dino-Tan}
  K.~Goulianos, Phys.\ Lett.\ B358 (1995) 379; \\
  C.I.~Tan, Phys.\ Rep.\ 315 (1999) 175.
\bibitem{Levin}
  E.~Gotsman, E.~Levin and U.~Maor, Phys.\ Rev.\ D60 (1999) 094011.
\bibitem{Martin}
  V.A.~Khoze, A.D.~Martin and M.G.~Ryskin, \verb=hep-ph/0007083=
\bibitem{pompyt}
  P.~Bruni and G.~Ingelman, DESY 93-187 (1993). \\
  \url{www3.tsl.uu.se/thep/pompyt/}
\bibitem{Herwig}
  G.~Marchesini et al., Comput.\ Phys.\ Commun.\ 67 (1992) 465; \\
  \verb=hep-ph/0011363=.
\bibitem{Pomwig}
  B.E.~Cox and J.R.~Forshaw, \verb=hep-ph/0010303=; \\
  \url{www.hep.man.ac.uk/pomwig/}
\bibitem{IS}
  G.~Ingelman and P.~Schlein, Phys.\ Lett.\ B152 (1985) 256.
\bibitem{h1pdf}
  H.~Jung and J.P.~Phillips, private communication.
\bibitem{DL}
  A.~Donnachie and P.V.~Landshoff, Phys.\ Lett.\ B191 (1987) 309; \\
  Nucl.\ Phys.\ B303 (1988) 634.
\bibitem{GRV}
  M.~Gl\"uck, E.~Reya and A.~Vogt, Z.~Phys.\ C67 (1995) 433.
\bibitem{pion-flux}
  K.~Golec-Biernat, et al., Phys.\ Rev.\ D56 (1997) 3955;\\
  A.~Szczurek, N.N.~Nikolaev and J.~Speth, Phys.\ Lett.\ B428 (1998) 383.
\bibitem{GRV-pion}
  M.~Gl\"uck, E.~Reya and A.~Vogt, Z.~Phys.\ C53 (1992) 651.
\bibitem{CFL}
  B.E.~Cox, J.R.~Forshaw and L.~L\"onnblad, JHEP (1999) 9910.
\bibitem{Pythia}
  T.~Sj\"ostrand, Comput.\ Phys.\ Commun., 82 (1994) 74.\\
  \url{www.thep.lu.se/~torbjorn/Pythia.html}
\bibitem{Pythia-MI}
  T.~Sj\"ostrand and M.~van Zijl, Phys.\ Rev.\ 36 (1987) 2019
\end{thebibliography}
\end{document}